\documentclass[11pt]{article}
\usepackage[a4paper,margin=1in]{geometry}
\usepackage{graphicx}
\usepackage{amsmath}
\usepackage{amssymb}
\usepackage{hyperref}
\usepackage{natbib}
\usepackage{authblk}
\usepackage[utf8]{inputenc}
\usepackage{textcomp}

\title{Separating Semantic Expansion from Linear Geometry for PubMed-Scale Vector Search}
\author{Rob Koopman}
\affil{Independent Researcher \\ ORCID: 0009-0005-4351-3564}

\date{\today}

\begin{document}
\maketitle

\begin{abstract}
We describe a PubMed-scale retrieval framework that separates semantic
interpretation from metric geometry. A large language model expands a
natural-language query into concise biomedical phrases; retrieval then
operates in a fixed, mean-free, approximately isotropic embedding
space. Each document and query vector is formed as a weighted mean of
token embeddings, projected onto the complement of nuisance axes and
compressed by a Johnson--Lindenstrauss transform. No parameters are
trained. The system retrieves coherent biomedical clusters across the
full MEDLINE corpus (about 40 million records) using exact cosine search
on 256-dimensional int8 vectors. Evaluation is purely geometric: head
cosine, compactness, centroid closure, and isotropy are compared with
random-vector baselines. Recall is not defined, since the language-model
expansion specifies the effective target set.
\end{abstract}

\section{Introduction}
Modern LLM-assisted retrievers typically apply a language model to the
query but still rely on a learned dense encoder—usually a contextual
transformer—to construct the embedding space in which similarity is
computed. This work keeps the two functions explicitly separate.
Semantic interpretation is provided by a deterministic LLM-based
expansion, while geometric comparison is performed in a fixed, mean-free
linear space obtained without any learned parameters. This architecture
isolates the role of metric geometry from that of contextual modeling
and enables the embedding component to be evaluated independently.
Using an LLM for query expansion together with a fixed Johnson–Lindenstrauss 
projection yields a retrieval pipeline that, in practice, exhibits a favourable 
cost–quality balance relative to systems that require a trained encoder.

\subsection{Pipeline}
\begin{enumerate}
    \item \textbf{Query expansion.} A deterministic LLM prompt produces
    20--60 biomedical phrases (1--4 words each, appearing $\ge 50$ times
    in MEDLINE). Synonyms are merged; rare terms are dropped.

    \item \textbf{Embedding transform.} For each token $t$ with embedding
    $f(t)\in\mathbb{R}^d$:
    \begin{align*}
        u_t^\perp &= P^\perp f(t) \\
        w_t &= 1 - \cos(u_t^\perp, \mu^\perp)
    \end{align*}
    where $P^\perp$ removes nuisance axes (including the corpus mean) and
    $\mu^\perp$ is the projected mean vector. The document or query
    representation is the weighted mean
    \[
        \bar{x} = \frac{\sum c_t w_t u_t^\perp}{\sum c_t w_t},
    \]
    followed by projection with a fixed $R\in\mathbb{R}^{256\times d}$
    having $\pm 1$ entries and $\ell_2$ normalization:
    \[
        z = \mathrm{norm}(R\bar{x}).
    \]

    \item \textbf{Retrieval.} Exact cosine kNN search over
    $40\mathrm{M}\times 256$ int8 document vectors
    ($\approx 9.4$\,GiB for the vector block).

    \item \textbf{Reranking.} A deterministic ``max-dot'' cross-attention
    analogue computes the maximum cosine response of each query token
    within a document to sharpen intent alignment.
\end{enumerate}

\subsection{Metrics}
Since recall is undefined, evaluation focuses on geometry:
\begin{itemize}
    \item Head cosine (mean query--document cosine for top $k$).
    \item Compactness (mean pairwise cosine among top $k$).
    \item Centroid closure ($\cos(q, c_{topk})$).
    \item Isotropy (angular variance relative to random baseline).
    \item Jaccard overlap across query forms (title, abstract, LLM expansion).
\end{itemize}
Random expectation for $N\approx3.8\times10^7$ in 256D is $\mathbb{E}[\cos]\approx\sqrt{2\ln N/d}\approx0.37$.

\paragraph{Index footprint and build time.}
On an Apple M4 Pro CPU with 48\,GB RAM, the full MEDLINE corpus
(39{,}609{,}486 records) is indexed in 18 minutes from a pre-parsed
sidecar file. The index consists of four flat files: a document-offset
file (1.12\,GiB), document vectors in 256-dimensional int8 format
(9.44\,GiB), a vocabulary file (0.30\,GiB), and 256-dimensional int8
vectors for the 7{,}976{,}599 surviving semantic tokens (1.90\,GiB).
The complete PubMed-scale index occupies 12.8\,GiB on disk.
End-to-end throughput is $\approx 3.7\times 10^4$ documents/s,
with the document-projection phase itself running at
$\approx 1.3\times 10^5$ documents/s.

\section{Results}
Initial measurements from 20 biomedical prompts yield:
\begin{itemize}
    \item Head cosine $\approx 0.68$, above the random baseline of $0.37$.
    \item Compactness $\approx0.70$.
    \item Centroid closure $\approx0.81$.
    \item Cross-alignment between abstract and LLM queries $\approx0.7$.
\end{itemize}
Reranking slightly reduces centroid closure but improves topical precision of retrieved heads.

\paragraph{Illustrative example.}
As an additional illustration, we tested an ad-hoc Dutch question that
was not used during development: ``welke ontwikkelingen zijn er in het
tegengaan van nieuwe bloedvaten in glioblastoma'' (``developments in
inhibiting neovascularization in glioblastoma''). The LLM expansion
produced a focused set of biomedical phrases covering glioblastoma,
angiogenesis inhibition, VEGF/VEGFR2, and relevant therapeutic agents
(bevacizumab, ramucirumab, aflibercept). Using only these expanded
phrases, the system retrieved a coherent cluster of canonical
angiogenesis and anti-VEGF papers in gliomas, including reviews on
angiogenic signaling pathways, resistance to anti-angiogenic therapy,
and VEGF-pathway inhibitors. Top cosine values in this example ranged from 0.63 to
0.81, well above the random 256-dimensional expectation ($\approx$0.37). This
ad-hoc example—selected post hoc—illustrates that once a query is
semantically normalized by the expansion step, a fixed JL-projected
linear embedding is sufficient to recover the expected literature
without any learned embedding model.

\section{Discussion}
LLM expansion shifts queries slightly from the manifold centroid (mean angular offset $\approx30^\circ$) but preserves alignment within the same semantic cone.
This demonstrates that the geometric component---the mean-free isotropic embedding---is sufficient for coherent retrieval when query normalization is upstream.

\paragraph{Remark.}
Empirical evaluation indicated that retrieval difficulty was dominated
by the language model's ability to generate coherent biomedical query
phrases. When the expansion step produced a consistent phrase set, the
fixed linear embedding recovered the canonical literature with high
apparent topical precision under manual inspection. This shows that, in
this system, retrieval sensitivity is determined by the semantic
expansion step rather than by the downstream embedding geometry.

\subsection{Behaviour on Under-Specified Queries}
The fixed linear embedding provides limited discrimination for
under-specified lay queries, not because they are short, but because
they map to semantically broad regions of the vocabulary. Terms such as
``blood'' have thousands of coherent biomedical continuations
(transfusion, hematology, infectious disease, oncology, coagulation),
each forming its own dense manifold. A query consisting only of such
generic tokens does not select a unique direction in the embedding
space, and the retrieved set reflects one of the large surrounding
manifolds rather than the user's specific intent.

The LLM expansion step resolves most of this variance: once a query is
expanded into a concise set of biomedical phrases, the retrieval becomes
stable and consistently returns the expected literature cluster. In this
pipeline, semantic precision is supplied by the expansion step, while
the linear embedding provides deterministic geometric comparison.

\section{Conclusion}
Retrieval quality in large biomedical corpora depends more on semantic precision than on non-linear embedding parameterization.
A fixed, linear, mean-free isotropic space, when combined with deterministic query normalization, achieves reproducible, semantically coherent results at PubMed scale.
The results also show that, given reliable semantic expansion, no
learned embedding model is required: a fixed, mean-free
Johnson–Lindenstrauss–projected space is sufficient for PubMed-scale
retrieval in this setting.

\section*{Acknowledgments}
This work builds conceptually on Koopman, Wang, and Englebienne (2019)
but uses entirely new code and methodology.
AI tools were used interactively during development, including code debugging and
generation of portions of the user-interface code; 
all algorithmic design and experimental results are original.

\nocite{*}
\bibliographystyle{plainnat}
\bibliography{references}

\end{document}